\documentclass[]{raa}            

\usepackage{booktabs,threeparttable}
\usepackage{amssymb}
\usepackage{multirow}
\usepackage{longtable}
\usepackage{txfonts}
\usepackage[authoryear]{natbib}
\usepackage{graphicx,times}             

\begin{document}

   \title{Optical Monitoring of the Seyfert Galaxy NGC 4151 and Possible Periodicities in the Historical Light Curve 
}

   \volnopage{Vol.0 (200x) No.0, 000--000}      
   \setcounter{page}{1}          

   \author{ Di-Fu Guo
      \inst{1},
      Shao-Ming Hu
      \inst{1},
      Jun Tao
      \inst{2},
      Hong-Xing Yin
      \inst{1},
       Xu Chen
      \inst{1}
       \and Hong-Jian Pan
      \inst{2}        }

   \institute{Shandong Provincial Key Laboratory of Optical Astronomy and Solar-Terrestrial Environment, Institute of Space Sciences, School of Space Science and Physics, Shandong University, Weihai, 264209, China; {\it husm@sdu.edu.cn}\\
           \and
             Shanghai Astronomical Observatory, CAS, 80 Nandan Road, Shanghai, 200030, China\\
   }

   \date{Received~~2009 month day; accepted~~2009~~month day}

\abstract{
We report $B$, $V$, and $R$ band CCD photometry of the
Seyfert galaxy NGC 4151 obtained with the 1.0~m telescope at Weihai Observatory of Shandong University and the 1.56 m
telescope at Shanghai Astronomical Observatory from 2005 December to
2013 February. Combining all available data from literature, we have constructed a historical light curve from 1910 to 2013 to
study the periodicity of the source using three different methods (the Jurkevich method, the Lomb-Scargle periodogram method
and the Discrete Correlation Function method). We find possible periods of  $P_1=4\pm0.1$, $P_{2}=7.5\pm0.3$ and $P_3=15.9\pm0.3$ yr.
\keywords{methods: data analysis -- galaxies: active -- galaxies: individual (NGC 4151)} }

   \authorrunning{ D. F. Guo et al. }            
   \titlerunning{Periodicities in NGC 4151}  

   \maketitle

%
%
\section{Introduction}           
\label{sect:intro}

The nature of active galactic nuclei (AGNs) is an important and open question in astrophysics.
Photometric observation of AGNs is an important tool for constructing their
light curves and for studying the variability behavior over different time
scales. Photometric observations have been made for a long
time, so it is possible to search for periodicity in the light curves
of a number of objects \citep[e.g.][]{Kidger92, Liu95, Fan97,
Fan98, Fan02a, Fan10, Qian04, Tao08, Lihuaizhen}. Based on the variability analysis,
we can get much information about the physical mechanisms of AGNs.
In addition, a confirmed periodicity would help us investigate the relevant
physical parameters and limit the physical models in AGNs \citep{Lainela99}.

NGC 4151 (Seyfert type 1.5) is one of the nearest
($z=0.00332$, $D=13.2$ Mpc when a Hubble constant of $75 \rm km
s^{-1}\rm Mpc^{-1}$ is used) and brightest Seyfert galaxies.
It is also one of the best studied objects across the entire
electromagnetic spectrum owing to its brightness and variability properties.
The nucleus of this galaxy shows flux variability at a wide range of wavelength,
with time scales from a few hours in the  hard X-ray \citep{Yaqoob93} to several months in the infrared \citep{Oknyanskij99}.

The photometric observations of NGC 4151 began as early as 1906. The nucleus of NGC 4151 was
first discovered in 1967 as variable in optical region \citep{Fitch67},
and the variability was confirmed by \cite{Zaitseva69}. Since then it has been intensively
monitored by several monitoring campaigns \citep[e.g.][]{Clavel90, Yaqoob93, Kaspi96}.
Optical variability time scales ranged from tens of minutes to
decades were reported by many investigators \citep{Lyutyi89, Lyutyi77, Lyutyj87, Longo96, Guo06, Oknyanskij07}. \cite{Oknyanskij12} pointed out that NGC 4151 has different variable components:
 fast variations with time scale about tens of days; slow variations with time scale about several years;
 very slow component with time scale about tens of years. NGC 4151 is one of the very few AGNs whose optical light curve data spans more than a century,
so it provides us a very good opportunity to analyze its periodicity.

In this work, we present optical ($BVR$) observations over
 7.2 years using the 1.0 m telescope at Weihai Observatory of Shandong University and 1.56 m
telescope at Shanghai Astronomical Observatory (SHAO). In section 2, we give a
description of our observations and data processing. Then in
sections 3 and 4, we present the light curves and the periodicity
analysis. Discussions and conclusions are given in section 5 .
\section{Observations and data reduction}
From 2005 December to 2013 February, 1587 observations were obtained on 22 nights using the 1.0m telescope
at Weihai Observatory of Shandong University ( observed from 2009 May to 2013 February) and the 1.56m
telescope at Sheshan Station of Shanghai Astronomical Observatory
( observed from 2005 December to 2008 February). The seeing at Weihai usually varies from 1.2$\arcsec$  to
2.5$\arcsec$. The one-meter Cassegrain telescope at Weihai Observatory was
equipped with a  back illuminated PIXIS 2048B CCD camera from the
Princeton Instruments company and a standard Johnson/Cousins set of
$UBVRI$ filters controlled by a dual layer filter wheel from
American Astronomical Consultants and Equipment Inc. (ACE). The PIXIS camera has
 $2048\times2048$ square pixels and the pixel size is 13.5$\mu$m. The scale of the
 image is about 0.35$\arcsec$ per pixel and the field of view is about 11.8$'$ $\times$ 11.8$'$.
 Readout noise and gain of this CCD detector is 3.64 electrons
and 1.65 electrons/ADU, respectively, for slow readout and low-noise output setup. The standard Johnson and Cousins
filters ($B$, $V$, and $R$) were used during our observations. For each observing night, twilight
sky flats were taken, several bias frames were taken at the beginning
of the observation. All data were processed by bias and flat-field correction.
The instrumentation and data reduction information for SHAO is the same as \cite{Guo06}.
The task APPHOT of the IRAF software package was used to do the photometry.
Comparison stars 2 and 3 taken from \cite{Penston71} were used for calibration.
An aperture radius about 13$\arcsec$ was adopted, which was larger enough
than the minimum aperture recommended by \cite{Cellone00}.
Then we were able to derive magnitude of the source by differential photometry.
The error is given as below
\begin{equation}\label{}
\sigma=\sqrt{(m_{2}-\overline{m})^{2}+(m_{3}-\overline{m})^{2}},
\end{equation}
where $m_{2}$ and $m_{3}$ is the magnitude of NGC 4151 calibrated by the $2$th and $3$th
comparison star respectively, whereas $\overline{m}$ is the mean magnitude of the
object obtained from the comparison stars.

\section{Light curves}

Panel a, b and c in Fig. 1 shows the light curves of NGC 4151 and the magnitude
difference between comparison stars 2 and 3 in $B$, $V$, and $R$ band, respectively. Different constants were added to the
corresponding differential magnitudes of the comparison stars for clarity. During our observations,
variations of 0.669 mag (12.199 to 12.868 ) in $B$ band, 0.964 mag (11.542 to 12.506 )
 in $V$ band, and 0.451 mag (10.875 to 11.326 ) in $R$ band were detected.
Our observations in $V$ band illustrates that NGC 4151 was decreasing in its
brightness from the beginning of our campaign and reached its minimum in 2008 February,
then it began to brighten and reached its maximum in 2011 April, finally it declined slowly in brightness again on the whole with some fluctuations.
The light curve can be roughly fitted by a sine function during our 7.2 years observation period,
 but more data are needed to confirm whether this periodicity is real.
The patterns of optical variability in these three bands are similar.

In order to detect the microvariability, both C test \citep{Jang97, Diego10} and F test \citep{Diego10, Gaur12} tests have been performed on those 10 out of 22 nights when we have more than five observations in the three bands simultaneously. However, no significant microvariability was detected during our observations.

\begin{figure*}
\includegraphics[angle=0,scale=0.7]{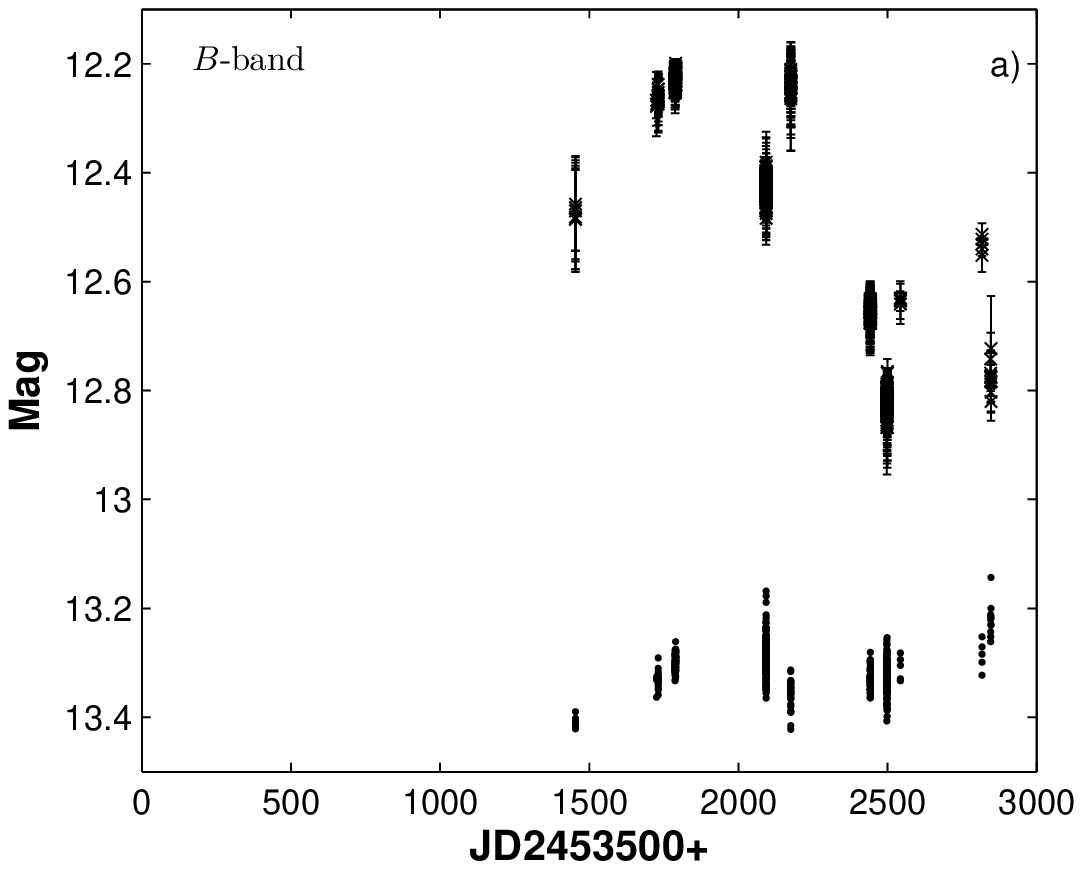}
\includegraphics[angle=0,scale=0.7]{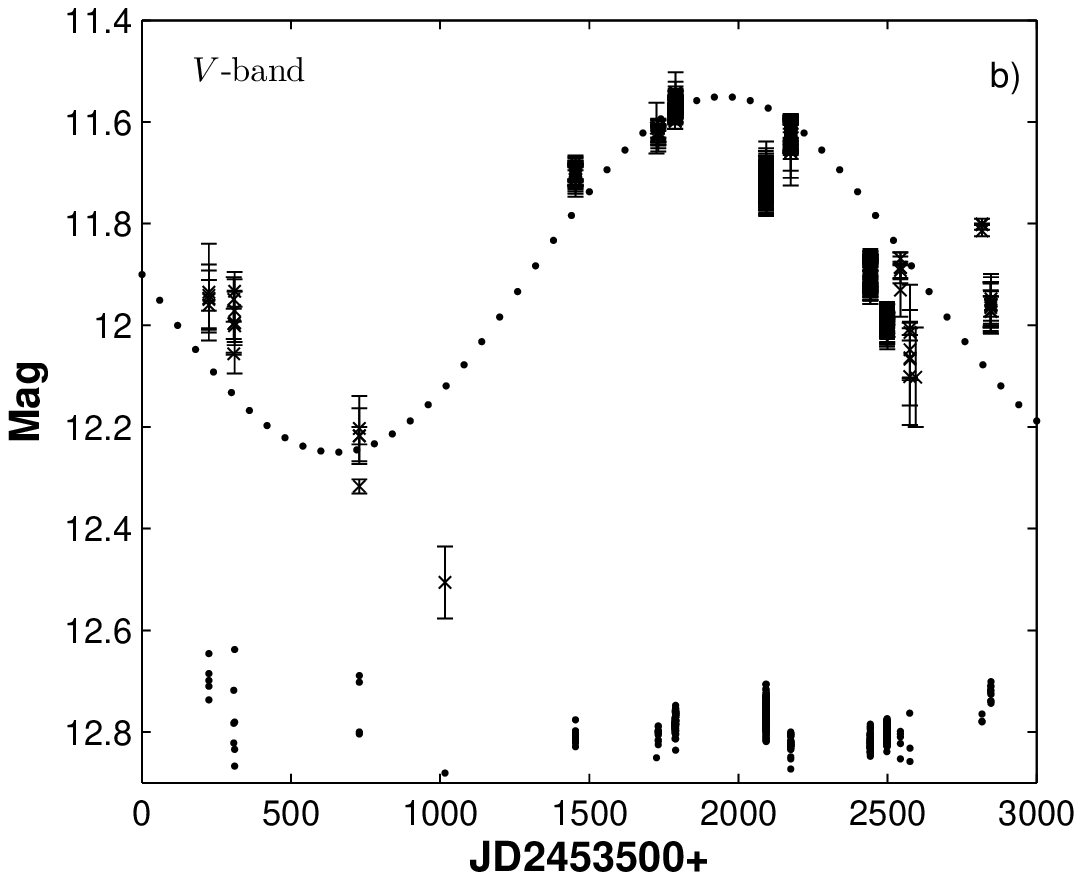}
\centering
\includegraphics[angle=0,scale=0.7]{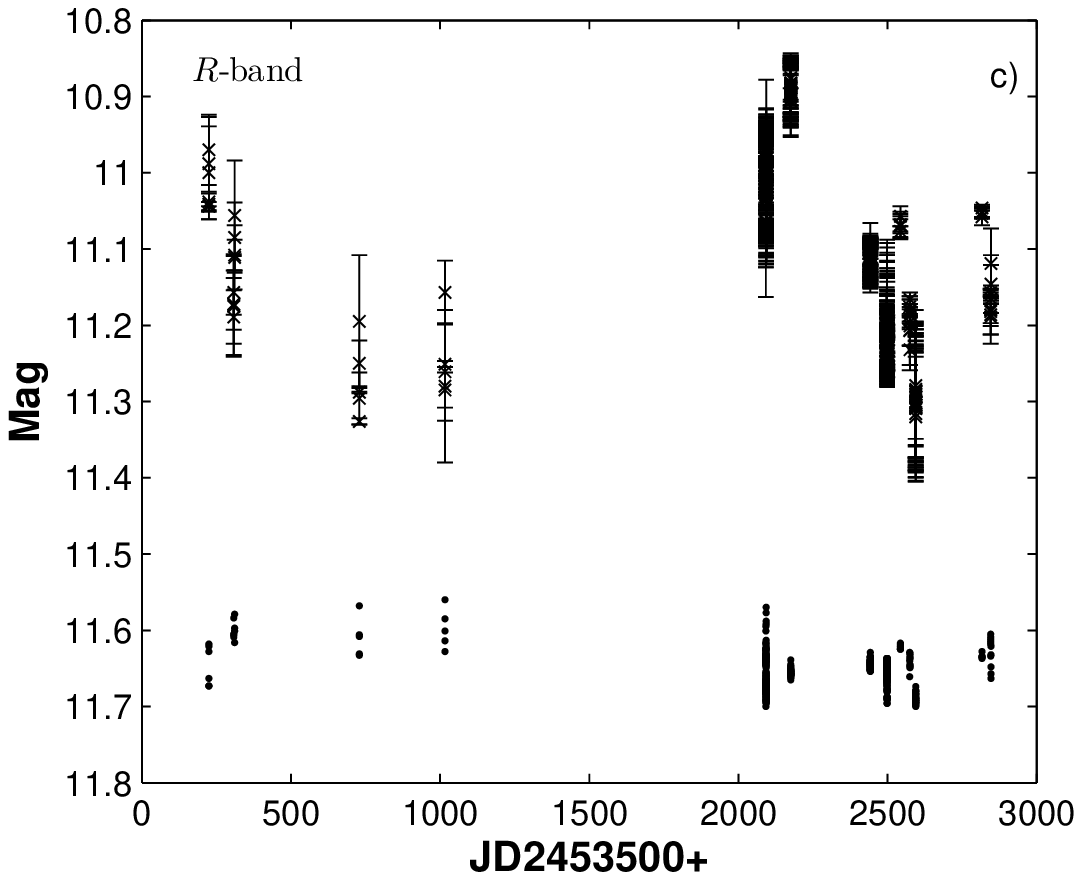}
\hfill \caption{Light curve of NGC 4151. x-marks denote $B, V$, and $R$ band Light curve of NGC 4151. Dots denote the magnitude difference of the comparison stars 2 and 3 (constants were added). Dotted line in panel b is the fitted sine function for $V$ band data.}
\label{}
\end{figure*}

\section{Periodicity Analysis}
Many authors have investigated the variability periodicity of NGC 4151.
\cite{Oknyanskij83} suggested a possible period of 16 yr using data from \cite{Pacholczyk71}.
\cite{Lyutyi87} found periods of 4 and 14 yr using the method of \cite{Deeming75} to a
set of about 400 photoelectric data. No evident periodicity was found by \cite{Longo96}.
\cite{Fan99} obtained a period of 14.08$\pm$0.8 yr using the Jurkevich method.
\cite{Oknyanskij07} found a periodic component about 15.6 years using the Fourier (CLEAN) algorithm.
 \cite{Oknyanskij12} suggested that NGC 4151 has different long-term variable time scales ranged from several years
 to tens of years,
 so we have reconstructed the historical light curve by combining our own observations with data in the literature
\citep{Zaitseva69, Lyutyi73, Lyutyi77, Belokon78, Lyutyi99,
Doroshenko01, Oknyanskij12, Roberts12} and the raw photographic data as early as 1910 to research periodicity.
The long-term light curve in $B$ band was shown in Fig. 2. The data used in this paper are mainly from \cite{Lyutyi99}
, \cite{Doroshenko01} and \cite{Oknyanskij12}. These data are consistent with each
other and they are from many published data. Their data set was called
Crimean series subsequently.
We found that the data derived from \cite{Roberts12} was
about 0.35 mag fainter compared with Crimean data which overlaid on the same day.
 Unfortunately, Our data didn't overlap with Crimean data set. However,
we noticed that for Crimean data set, the brightness of NGC 4151 remained nearly
constant between JD 2456042 and JD 2456044, and we just had observations on JD 2456043.
Further more, no microvariability was detected during our observations.
So a linear interpolation can be used to combine our data with the Crimean data. A constant about 0.25 mag was subtracted in order to combine our data with theirs.

In order to give more uniform weighting to different epochs,
we carried out a 10-day averaging for all the available data. This bin size is short enough compared with the long-term periods which we are looking for (years) and thus unlikely to distort long-term variations.

\begin{figure*}
\centering
\includegraphics[angle=0,scale=0.8]{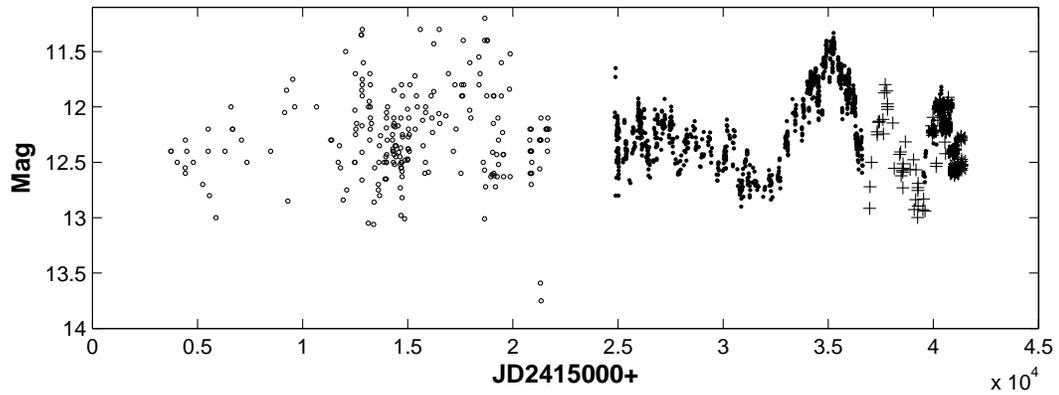}
\hfill \caption{Historical light curve of NGC 4151 from 1910 to 2013 in $B$ band.
  Dots denote the Crimean series data, circles denote the raw photographic data
  , pluses denote the data from \cite{Roberts12} and stars denote
  our observations. \label{}}
\end{figure*}

\subsection{Jurkevich Method}

\begin{figure*}
\includegraphics[angle=0,scale=0.60]{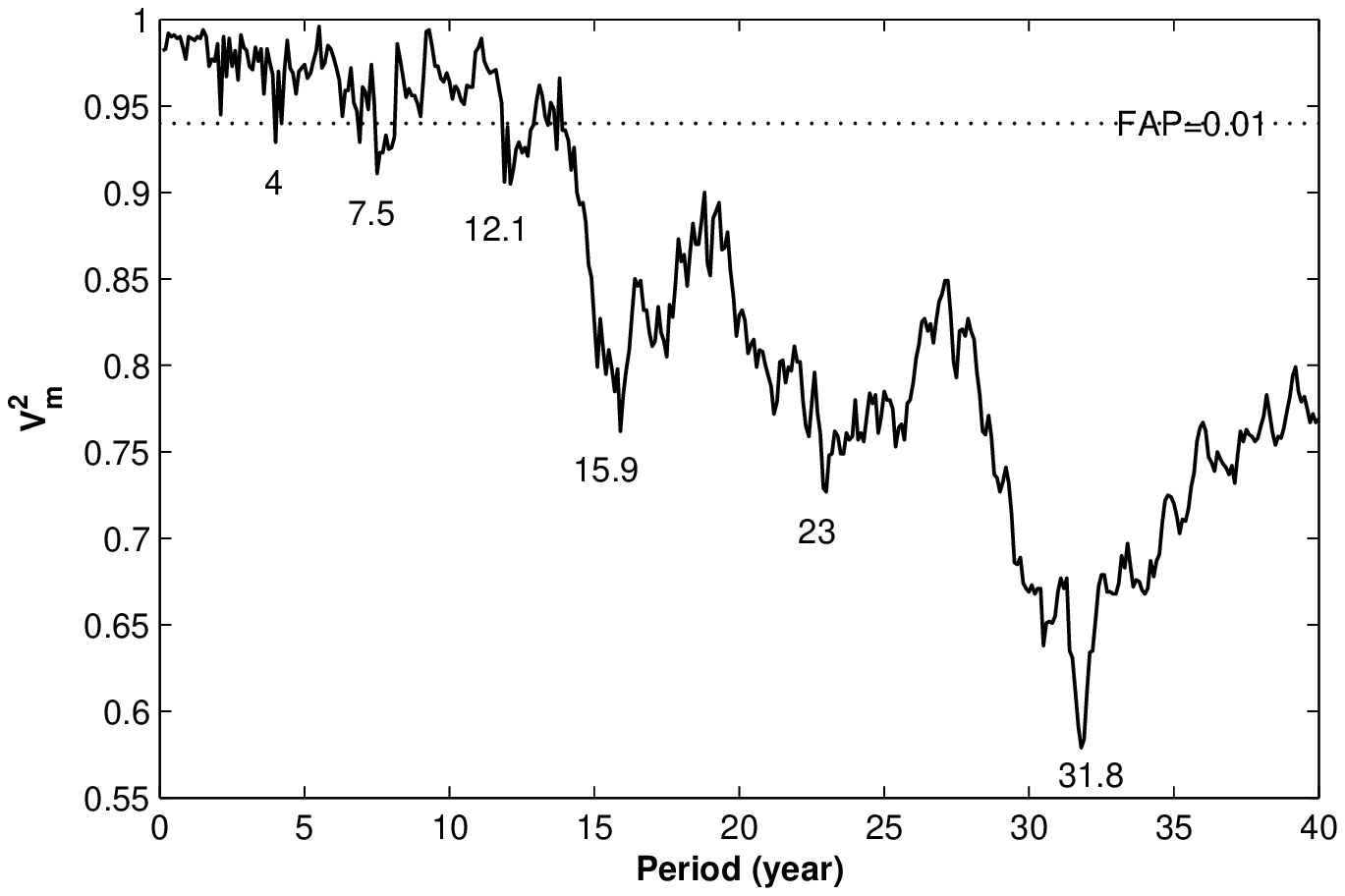}
\includegraphics[angle=0,scale=0.58]{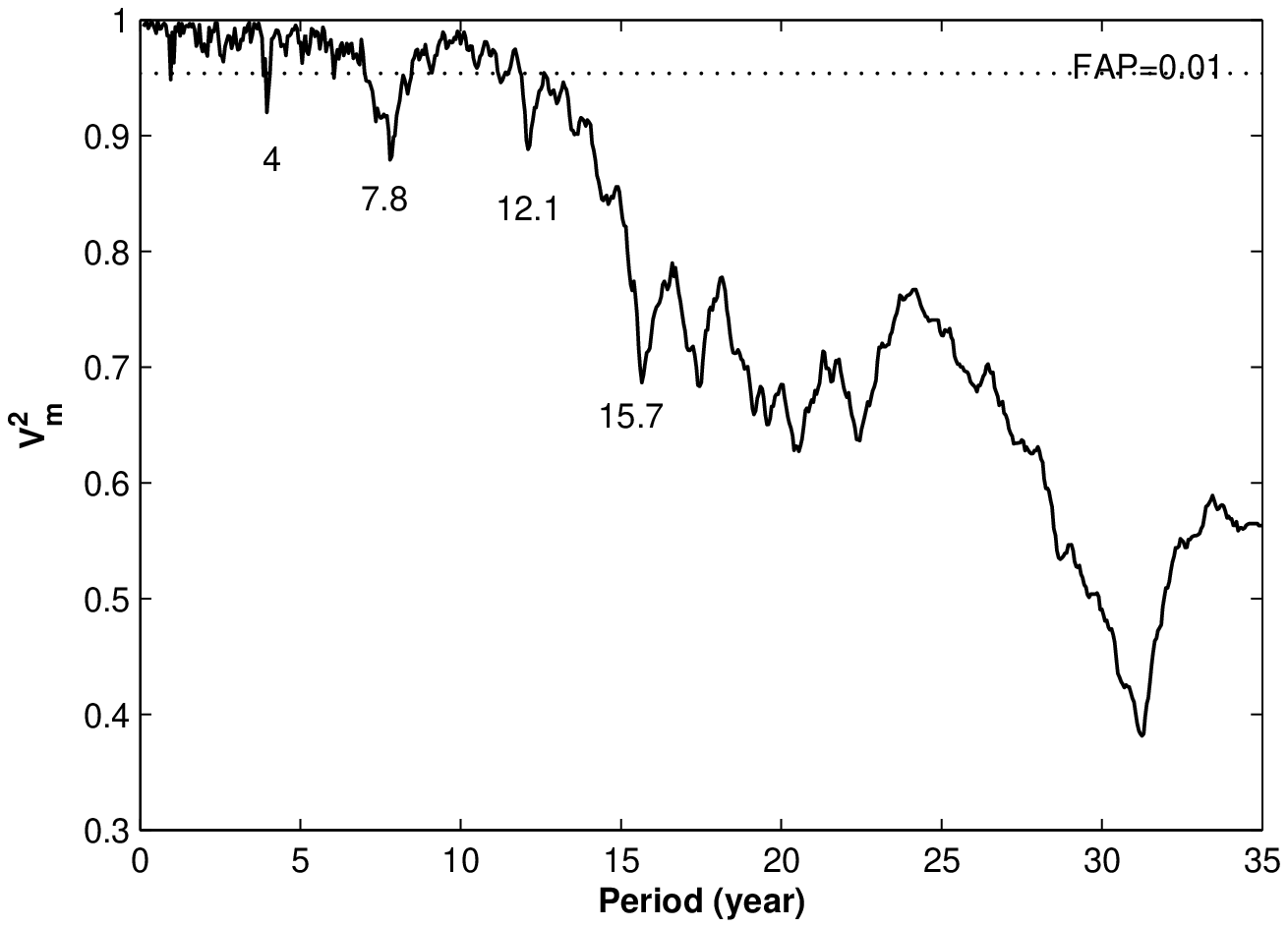}
\hfill \caption{Relationship between the trial period and $V_{m}^{2}$. Plot in the left panel and right panel was derived by using
  all the available data and the post-1968 data, respectively.}
\label{}
\end{figure*}

The Jurkevich method (JV)\citep{Jurkevich71} is based on the expected mean square
deviation. It does not require an equally spaced observations, so it is less
inclined to generate a spurious periodicity compared to a Fourier analysis.
It tests a series of trial periods and the data are
folded according to the trial periods. Then all data are divided
into $m$ groups according to their phases around each trial period. The variance $V_{i}^{2}$
for each group and the sum of each group variance $V_{m}^{2}$  are
computed. If a trial period equals to the real one, then $V_{m}^{2}$
would reach its minimum. The detailed computation of the variances
is described in \cite{Jurkevich71}. In order to
estimate the reliability of the period, \cite{Kidger92}
introduced a parameter,
\begin{equation}\label{}
 f=\frac{1-V^{2}_{m}}{ V^{2}_{m}},
       \end{equation}\label{}
where $V^{2}_{m}$ is a normalized value. In the normalized plot,
 a value of $V^{2}_{m}=1.0$ means that $f=0$, hence there is
 no periodicity at al. The possible periods can be easily obtained from the plot.
 In general, if $f \geq 0.5$, it suggests that there is a strong periodicity in the data;
 if $f\leq 0.25$, usually indicates that the periodicity, if genuine, is
 a weak one. A further test is the relationship between the depth of the minimum and
 the noise in the \textquotedblleft flat\textquotedblright  section of the $V^{2}_{m}$ curve close to the detected
 period \citep{Kidger92, Fan06}. \cite {Fan10} pointed out that this method is not
 good enough to give a quantitative criterion and the False Alarm Probability \citep[FAP,][]{Horne86}
 can deal with all kinds of periodicity analysis methods if the variations (mainly) consist of
 randomly distributed noise. So we adopted FAP to give a quantitative criterion \citep[see][]{Fan10, Chen13} for the detected periods.
One should also keep in mind the confidence level derived by this method tends to be overestimated. Because the independent random values were used to evaluate the statistical significance of periods and it may neglected the fact that the dependent random values existed in the light curves of AGNs.

The results by the Jurkevich method with  m=5 are shown in the left panel of Fig. 3. and there are no
significant differences when m=10 was used. FAP level of 0.01 is also marked in the figure derived from Monte Carlo method as proposed by \cite {Fan10}.
From the figure, there are several obvious minimum values for $V^{2}_{m}$ whose FAP is smaller than 0.01,
indicating possible periods in the trial. The first minimum of $V^{2}_{m}=0.92$
is at the period $P_{1}=4\pm0.1$ yr. it is consistent
with the result of 4 yr found by  \cite{Lyutyi87}. The second minimum of $V^{2}_{m}=0.91$ is corresponding to
$P_{2}=7.5\pm0.3$ yr. The third minimum of $V^{2}_{m}=0.9$ is corresponding to $P_{3}=12.1\pm0.2$ yr. The fourth minimum of
 $V^{2}_{m}=0.76$ is corresponding to $P_{4}=15.9\pm0.3$ yr. The fifth minimum of $V^{2}_{m}=0.72$ is corresponding to $P_{5}=23\pm0.3$ yr
 and the sixth minimum of $V^{2}_{m}=0.58$ is corresponding to $P_{6}=31.8\pm0.2$ yr. We note that those periods have the following simple
 relationships: $P_{3}\approx 3P_{1}$, $P_{5}\approx 3P_{2}$ and $P_{6}\approx 2P_{4}$.

From Fig. 2, one can see a large gap between JD. 2436717 and JD. 2439849
 which may strongly affect the sensitivity of the method and result in the appearance
  of false period. So we apply both the Jurkevich method and the Monte Carlo method
  to the more homogeneous data set after 1968 (beginning from JD. 2439849, called post-1968 data hereafter) to calculate the periodicity and FAP (0.01). The results are shown in the right panel of Fig. 3.
One can see from the right panel of Fig. 3 that there are several obvious minimum values for $V^{2}_{m}$ whose FAP is smaller than 0.01. Taking the shorter time span (only about 45 years observation) which was less than six times of the period \citep{Kidger92} into consideration,
the periods larger than ten years were ruled out. The periods of
$4\pm0.1$ and $7.8\pm0.4$ yr found in the right panel are in good agreement with the periods of $P_{1}=4\pm0.1$
and $P_{2}=7.5\pm0.3$ yr found in the left panel of Fig. 3, respectively.

\subsection{Lomb-Scargle Periodogram Method}

\begin{figure*}
\includegraphics[angle=0,scale=0.6]{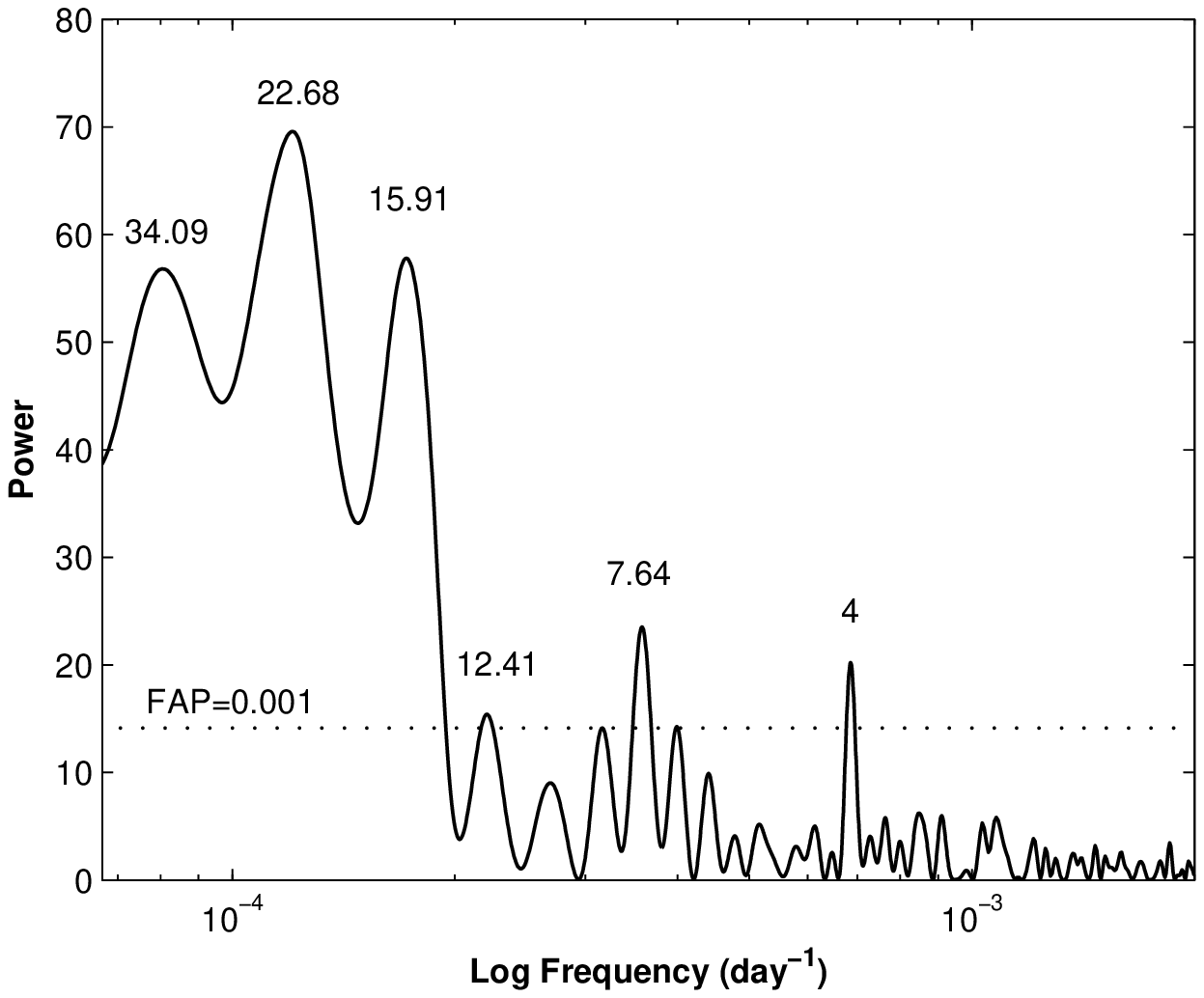}
\includegraphics[angle=0,scale=0.6]{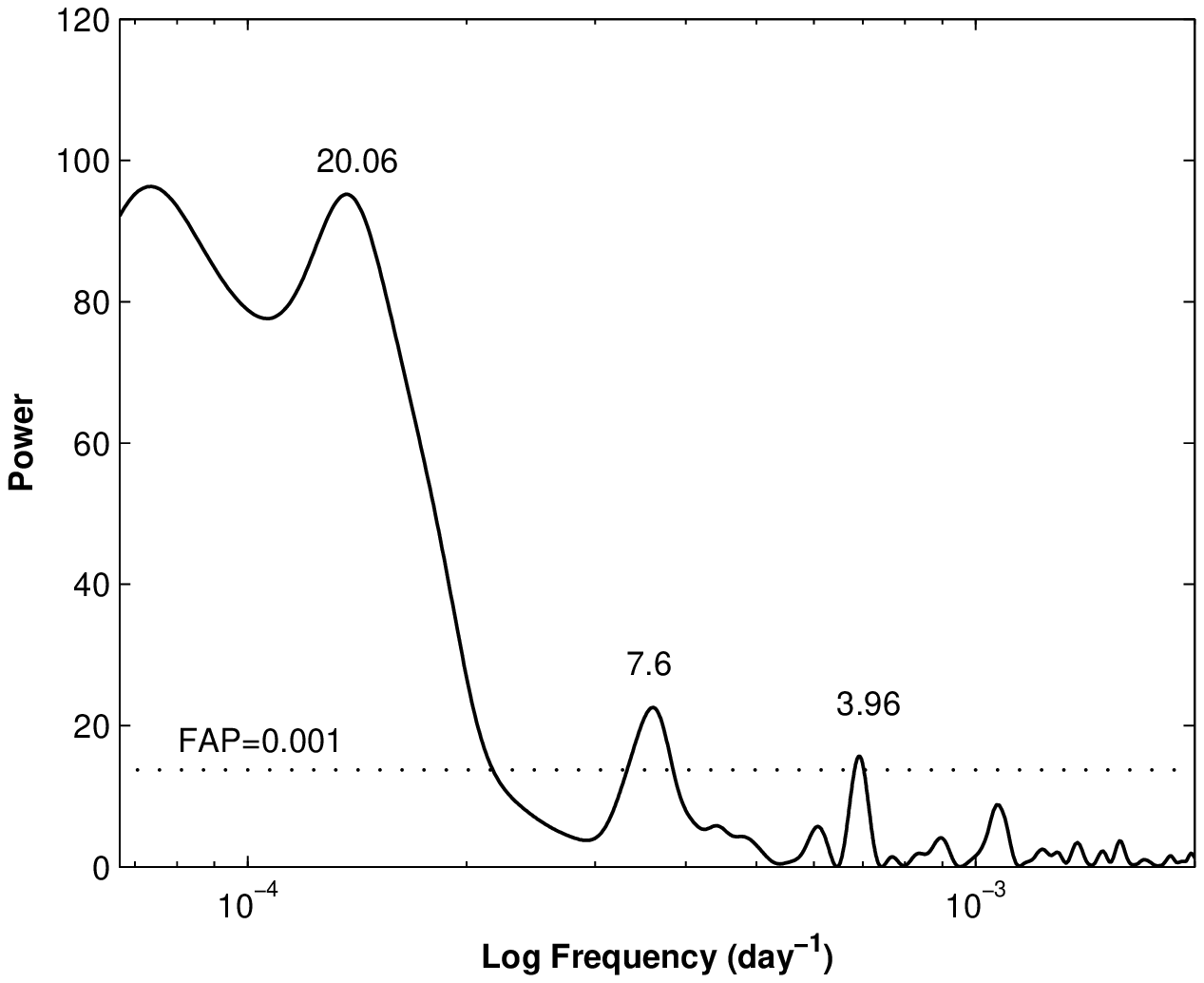}
\hfill \caption{Results of Lomb-Scargle method for NGC 4151. Power curve in the left panel and right panel was calculated using
  all the available data and the post-1968 data, respectively. Dotted lines denote the false alarm probability of 0.001.}
\label{}
\end{figure*}

In order to investigate the reliability of these results, we also
 performed the Lomb-Scargle normalized periodogram (LS) to analyze the periodicity.
The Lomb-Scargle normalized periodogram \citep{Lomb76, Scargle82, Press94}
is a powerful method which can be applied to the periodicity analysis of
random and unevenly sampled observations.

 For  a time series $X(t_k)(k=0,1....,N_0)$, the periodogram as a function of frequency $\omega$,
 is defined as

 \begin{equation}\label{}
 P_x(\omega)=\frac{1}{2}\{\frac{[\Sigma_{i}X(t_i)\cos\omega(t_i-\tau)]^2}{\Sigma_{i}\cos^2\omega(t_i-\tau)}+\frac{[\Sigma_{i}X(t_i)\sin\omega(t_i-\tau)]^2}{\Sigma_{i}\sin^2\omega(t_i-\tau)}\},
 \end{equation}\label{}
where $\tau$ is defined by the formula
  \begin{equation}\label{}
  \tau=\frac{1}{2\omega}\tan^{-1}[\frac{[\Sigma_{i}\sin{2\omega{t_i}}}      {\Sigma_{i}\cos{2\omega{t_i}}}].
 \end{equation}\label{}
Here and throughout, $\omega$ is the angular frequency and $\omega=2\pi\nu$, so the periodogram is also a function of the frequency $\nu$.
According to the definition of $P_x(\omega)$, the power in $P_x(\omega)$ would follow an exponential
 probability distribution if the signal $X(t_{k})$ is purely noise. This exponential distribution provides a convenient
 estimate for the probability that a given peak is a real signal, or it is only the result of randomly
 distributed noise. For a power level z, FAP is calculated as \citep [see][]{ Scargle82, Press94}
   \begin{equation}\label{}
  p(>z)\approx N\cdot exp(-z),
 \end{equation}\label{}
where N is the number of frequencies searched for the maximum. N is very nearly equal to the number of data points $N_0$ when the data points are approximately equally spaced and the estimate of N needs not be very accurate \citep{Press94}, and the half width at half maximum (HWHM) of the peak can be used to estimate the corresponding periodic error.

 The results of Lomb-Scargle Periodogram and false alarm probability are shown in Fig. 4. Power curves in the left panel and in the right panel were calculated using
  all the available data and the post-1968 data, respectively.
 From the left panel of Fig. 4, we could find several peaks ($P_1=4\pm 0.2$, $P_2=7.64\pm 0.3$, $P_3=12.41\pm 0.6$, $P_4=15.91\pm 0.9$, $P_5=22.68\pm 1.6$, $P_6=34.09\pm 3.2$) whose false alarm probability are smaller than 0.001.
For the results of $P_2$, $P_3$, $P_4$, $P_5$, $P_6$, we note that $P_5 \approx 3P_2$, $P_4 \approx 4P_1$,
$P_3 \approx 3P_1$. The periods found by the Lomb-Scargle Periodogram are in good agreement with the results found by the Jurkevich method using all the available data. We also apply the Lomb-Scargle Periodogram to the post-1968 data to calculate the periodicity and the results are shown in the right panel of Fig. 4. The periods of $3.96\pm0.1$ and $7.6\pm0.4$ yr is corresponding to $P_{1}=4\pm0.1$ and $P_{2}=7.5\pm0.3$ yr found by Jurkevich method using the same data set, respectively.

\subsection{Discrete Correlation Function Method}

\begin{figure*}
\includegraphics[angle=0,scale=0.55]{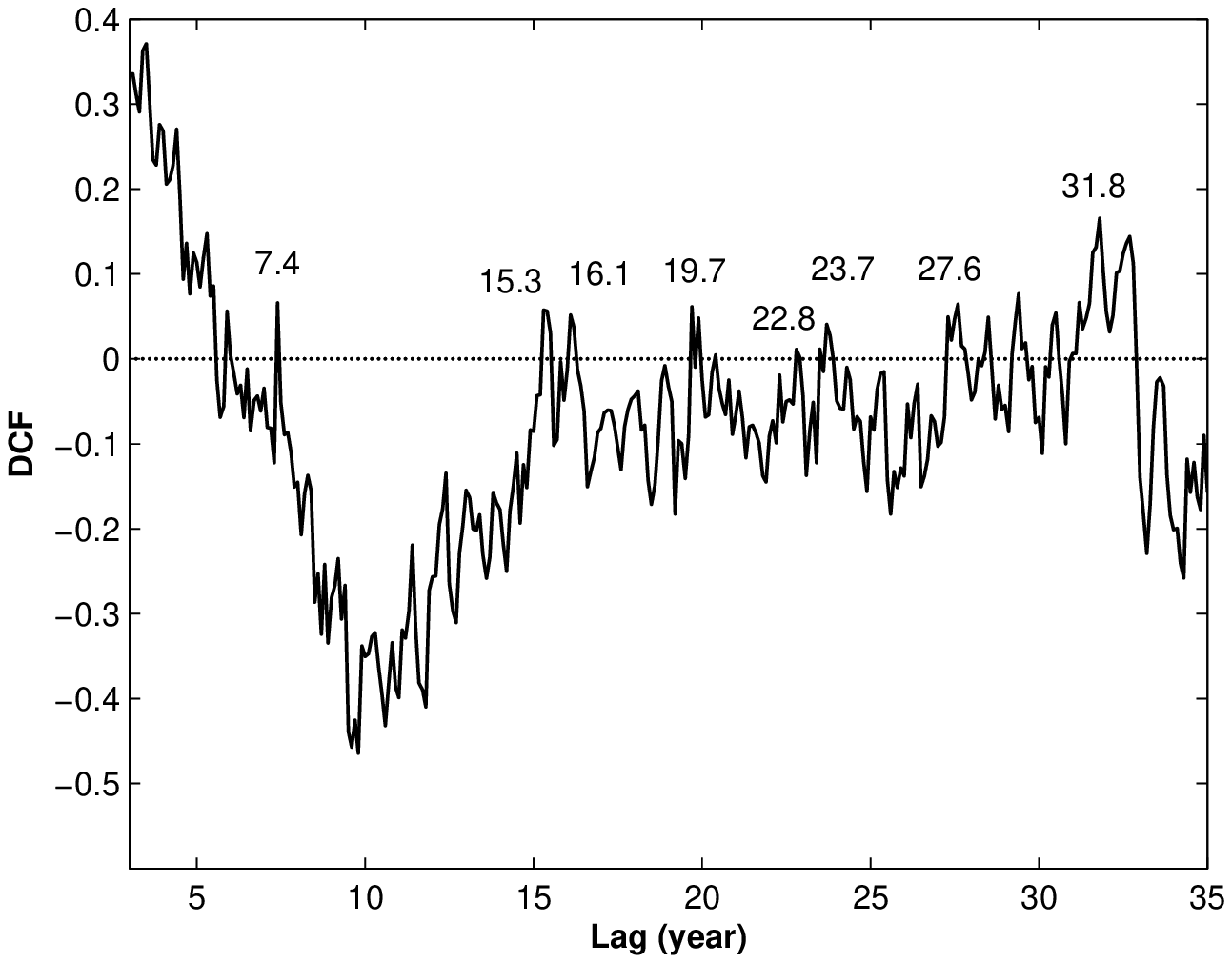}
\includegraphics[angle=0,scale=0.50]{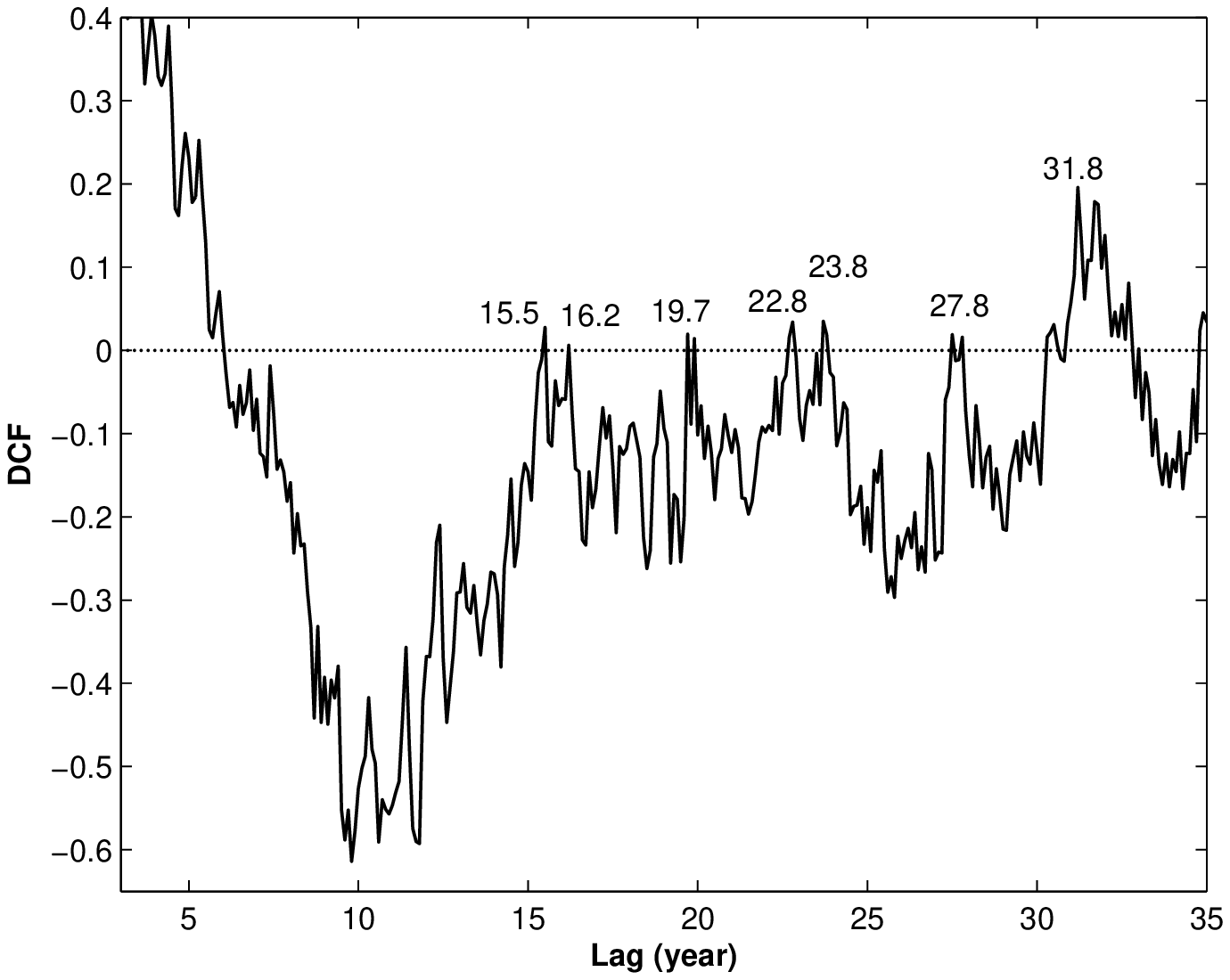}
\hfill \caption{Results of the Discrete Correlation Function method for NGC 4151. DCF curve in the left panel and right panel was obtained using
all the available data and the post-1968 data, respectively.}
\label{fit1}
\end{figure*}

 For comparing and further investigating  the reliability of these periods, we have
 performed the Discrete Correlation Function (DCF) method to analyze
 the periodicity. The DCF method, introduced by \cite{Edelson88},
 can be used to analyze unevenly sample variability data, with the
 advantage of no interpolation of data. It can be done as follows.
 Firstly, we calculate the set of unbinned discrete correlation function (UDCF) of all measured
 pairs($a_i,b_j$), i.e.
 \begin{equation}\label{}
 UDCF_{ij}=\frac{(a_i-\overline{a})(b_j-\overline{b})}{ \sigma_{a}\sigma_{b}},
 \end{equation}\label{}
where $\overline{a}$ and $\overline{b}$ are the average values of the data sets,
 $\sigma_{a}$ and  $\sigma_{b}$ are the corresponding standard deviations.
 Secondly, We averaged the points sharing the same time lag $\tau$ by binning
 the $UDCF_{ij}$ in the suitable sized time-bins, then $DCF(\tau)$ can be
 calculated as following,
 \begin{equation}\label{}
 DCF(\tau)=\frac{1}{M}\sum{UDCF_{ij}(\tau)},
 \end{equation}\label{}
where M is the number of pairs satisfying $\tau-\Delta{\tau}/2\leq\Delta{t_{ij}}<\tau+\Delta{\tau}/2$.
 Then, the standard error for each time lag $\tau$  is
 \begin{equation}\label{}
 \Delta{\tau}=\frac{1}{M-1}\{\sum[UDCF_{ij}-DCF(\tau)]^2\}^{0.5}.
 \end{equation}\label{}
In general, a positive peak constitutes a possible period and
the height of the peak indicates the periodicity strength \citep{Hufnagel92}.
The results of DCF are displayed in Fig. 5. DCF curve in the left panel and right panel was obtained using
all the available data and the post-1968 data, respectively. Several possible periods, estimated from positive peaks in the DCF,
are $7.4\pm0.1, 15.3\pm0.2, 16.1\pm0.2, 19.7\pm0.3, 22.8\pm0.3, 23.7\pm0.3, 27.6\pm0.3$ and $31.8\pm0.2$ yr for the left panel of Fig. 5;
$15.5\pm0.2, 16.2\pm0.3, 19.7\pm0.3, 22.8\pm0.3, 23.8\pm0.4, 27.8\pm0.4$ and $31.8\pm0.2$ yr for the right panel.
One can note that the periods found by DCF methods have multiple frequency relationships with the periods of $P_1=4\pm0.1$, $P_{2}=7.5\pm0.3$ or $P_3=15.9\pm0.3$ yr found
by Jurkevich method.

\subsection{Results}
From the analyses of the results given above, we summarise the main periods in Table 1.
For the periods of $P_1$, $P_2$, $P_3$, $P_4$, $P_5$ and $P_6$ discerned by the Jurkevich method and the Lomb-Scargle Periodogram (see Table 1), one can note that there exists a simple relationship: $P_3\approx4P_1$, $P_4\approx3P_1$,
 $P_5\approx3P_2$ and $P_6\approx3P_3$. According to the result by \cite{Kidger92},
 the period of $P_1=4\pm0.1$ yr found by the Jurkevich method is a week one, whereas $P_3=15.9\pm0.3$ yr is strong periodicity. At the same time, \cite{Oknyanskij07}, \cite{Bon12} also found a possible period of around 16 yr using different methods and data,
 so we think that the period $P_1=4\pm0.1$ and $P_3=15.9\pm0.3$ yr have different origin,
 although they have an astronomical multiple frequency relationship $P_3\approx4P_1$.
For a further comparison, we used the DCF method for periodicity analysis. However, there is no sign of 4
yr period in the DCF analysis, this may be due to that the period of $P_1$ is a weak one. Further more
the DCF analysis will dilute the sign of other periods if more than one period is present \citep{Fan06}. Although there is no sign of 4 yr period in the results of DCF method, it is interesting to note that the periods of $19.7\pm0.3,23.7\pm0.3$ and $27.6\pm0.3$ yr are about 5, 6, and 7 times of the $4$ yr (see Fig. 5), respectively. It implies that they have an astronomical multiple frequency relationship. Based on the analysis above, we think that
the period of $4\pm0.1$ yr is existent indeed. Taking into account that our light curve in $V$ band can be roughly fitted by a 7.1-year sine function, which is consistent with the period of $P_{2}=7.5\pm0.3$ yr,
 so the possible periods found by the three methods are $P_1=4\pm0.1$, $P_{2}=7.5\pm0.3$ and $P_3=15.9\pm0.3$ yr.

\begin{table*}
  \caption[]{Periodicity Analysis Results}
  \label{Tab:poptical}
  \begin{center}\begin{tabular}{llllllll}
  \hline\noalign{\smallskip}
                          & $P_1(yr)$         &  $P_2(yr)$         &  $P_3(yr)$                & $P_4(yr)$       & $P_5(yr)$      & $P_6(yr)$      &  \\   \hline
      JV$^{\diamondsuit}$   & $4\pm0.1$   &    $7.5\pm0.3$  &  $15.9\pm0.3$  & $12.1\pm0.2$ ($3P_1$ )   & $ 23\pm0.3$ ($3P_2$ )  & $31.8\pm0.2$ ($2P_3$)    &   \\
      JV(post-1968)$^*$   & $4\pm0.1$         &    $7.8\pm0.4$       &          &            &          &           \\
      LS$^{\diamondsuit}$  & $4\pm0.2$   &    $7.64\pm0.3$  &  $15.91\pm0.9$   & $12.1\pm0.6$ ($3P_1$ )  & $ 22.68\pm1.6$ ($3P_2$ )  & $34.09\pm3.2$ ($2P_3$) &   \\
      LS (post-1968)$^*$   & $3.96\pm0.1$    &    $7.6\pm0.4$        &     &         &             &        &            \\
      DCF$^{\diamondsuit}$    &    &        $7.4\pm0.1$     &  $16.1\pm0.2$        &        & $22.8\pm0.3$ ($3P_2$)  & $31.8\pm0.2$ ($2P_3$)        &           \\
      DCF (post-1968)$^*$   &      &          &  $16.2\pm0.3$          &            & $ 22.8\pm0.2$ ($3P_2$ )   & $31.8\pm0.2$ ($2P_3$)          &          \\
\hline
  \end{tabular}\end{center}
  \begin{tablenotes}
  \item[*] $\diamondsuit$ Using all the available data.
  \item[*]* Using photoelectric and CCD observational data after 1968.
 \end{tablenotes}
\end{table*}
\section{DISCUSSION AND CONCLUSIONS}

The variability mechanism of AGNs is not yet well understood and some models
have been proposed to explain the possible optical long-term periodic variations:
the binary black hole model, the disk-instability model and the perturbation model.
As for NGC 4151, \cite{Aretxaga94} found that the long term variability
of NGC 4151 can be well described by the starburst model. However, it had difficulty
in explaining the short timescale of the optical variability and the extreme energetics
\citep{Gopal-Krishna00}. \cite{Fan02b} adopted the structure function to $B$ band
data to discuss the emission origin in the Seyfert galaxy NGC 4151
and the result favored the disk instability model. \cite{Czerny03} concluded that the long time scale variability may be caused
by radiation pressure instability in the accretion disc by analysis
of the 90 yr of the optical data and 27 yr of the X-ray data using the
normalized power spectrum density (NPSD) and the structure function.
 \cite{Lyuty05} found that the pattern
of ultraviolet and optical variability in NGC 4151 agreed excellently with the theory
of disk accretion instability for a supermassive black hole suggested by \cite{Shakura73}.
\cite{Oknyanskij07} thought the 14-16 years circles
seen in the light curve probably correspond to some accretion dynamic time.
Analysis by \cite{Bon12} showed that periodic variations in the light curve and radial velocity curve
can be accounted for an eccentric, sub-parsec Keplerian orbit of a 15.9 yr period.


 The historical light curve of NGC 4151 showed strong variability and three
 possible periods were found by different periodicity analysis methods.
 The periods of $P_1=4\pm0.1$, $P_{2}=7.5\pm0.3$ and $P_3=15.9\pm0.3$ yr are
 consistent with the findings by \cite{Oknyanskij12} that NGC 4151 has different long-term variable time scales ranged from several to tens of years.
 The period of $P_1=4\pm0.1$ yr is in good agreement with the result found by \cite{Lyutyi87}, and the period of $P_3=15.9\pm0.3$ yr is also consistent
 with the results found by many investigators \citep{Oknyanskij83,Guo06,Oknyanskij07,Bon12}.
 The multiple periods we derived may imply the instabilities in the disk.

In our monitoring program, we have observed NGC 4151 from
2005 December to 2013 February. The observations clearly show that
the source is variable in the optical band with the variation
amplitudes of 0.669 mag, 0.964 mag and 0.451 mag in $B$, $V$, and $R$ bands, respectively.
The $B$ band historical light curve is constructed, which has a time span of
103 yr. Possible periods of $P_1=4\pm0.1$, $P_{2}=7.5\pm0.3$ and $P_3=15.9\pm0.3$ yr were found in the light curve by
 adopting the Jurkvech method, the Lomb-Scargle Periodogram method and the DCF method.

\begin{acknowledgements}
We thank the anonymous referee for his/her constructive comments and suggestions. This work is partially supported by the NSFC (Nos. 11203016, 11143012, 10778619, 10778701, 10903005), and by the NSF of Shandong Province (No. ZR2012AQ008).
\end{acknowledgements}

\label{lastpage}

\end{document}